# Iterative phase retrieval for digital holography

TATIANA LATYCHEVSKAIA,[1,2]

[1]*Physics Institute, University of Zurich, Winterthurerstrasse 190, 8057 Zurich, Switzerland*
[2]*Paul Scherrer Institut, Forschungsstrasse 111, 5232 Villigen, Switzerland*
*\*tatiana@physik.uzh.ch*

**Abstract:** This paper provides a tutorial of iterative phase retrieval algorithms based on the Gerchberg–Saxton (GS) algorithm applied in digital holography. In addition, a novel GS-based algorithm that allows reconstruction of 3D samples is demonstrated. The GS-based algorithms recover complex-valued wavefront-by-wavefront back-and forth propagation between two planes with constraints superimposed in these two planes. Iterative phase retrieval allows quantitatively correct and twin-image-free reconstructions of object amplitude and phase distributions from its in-line hologram. The present work derives the quantitative criteria on how many holograms are required to reconstruct a complex-valued object distribution, be it a 2D or 3D sample. It is shown that for a sample that can be approximated as a 2D sample, a single-shot in-line hologram is sufficient to reconstruct the absorption and phase distributions of the sample. Previously, the GS-based algorithms have been successfully employed to reconstruct samples that are limited to a 2D plane. However, realistic physical objects always have some finite thickness and therefore are 3D rather than 2D objects. This study demonstrates that 3D samples, including 3D phase objects, can be reconstructed from two or more holograms. It is shown that in principle, two holograms are sufficient to recover the entire wavefront diffracted by a 3D sample distribution. In this method, the reconstruction is performed by applying iterative phase retrieval between the planes where intensity was measured. The recovered complex-valued wavefront is then propagated back to the sample planes, thus reconstructing the 3D distribution of the sample. This method can be applied for 3D samples such as 3D distribution of particles, thick biological samples, and other 3D phase objects. Examples of reconstructions of 3D objects, including phase objects, are provided. Resolution enhancement obtained by iterative extrapolation of holograms is also discussed.

## 1. INTRODUCTION

In 1947, Dennis Gabor invented holography when he was working on improving the resolution of the recently invented electron microscope [1, 2]. Despite the relatively short electron wavelength of only a few pm, which is hundreds of times larger than the distances between atoms, the images acquired in electron microscope did not exhibit atomic resolution. The reason was the aberrations of the electron lens system [3]. Gabor's solution to the problem was truly original. He suggested removing all the lenses between the sample and the detector. In this arrangement, Gabor argued, the electron wave passes through the sample and part of the wave interacts with the sample; the scattered wave and the unscattered waves interfere on a distant detector thus forming a unique interference pattern that contains the complete information about the sample distribution. This principle is illustrated in Fig. 1(a). The original experimental arrangement proposed by Gabor is called Gabor type holography, or in-line holography, since the reference and the object wave share the same optical axis. In 1952, Haine and Mulvey acquired the first experimental electron in-line hologram and reconstructed it by optical means [4].

Shortly after Gabor published his first paper about the holographic principle, he published a longer follow-up paper where he described the presence of the "twin image" [5]. For in-line holography with spherical waves, the twin image is positioned centro-symmetrically to the original object relative to the point source, as illustrated in Fig. 1(b). For in-line holography with plane waves, the twin image is positioned symmetrically towards the hologram plane as shown in Fig. 1(c)-(d). In both situations, spherical or plane waves, the twin images are superimposed onto the reconstructed object and contaminate the object distribution.

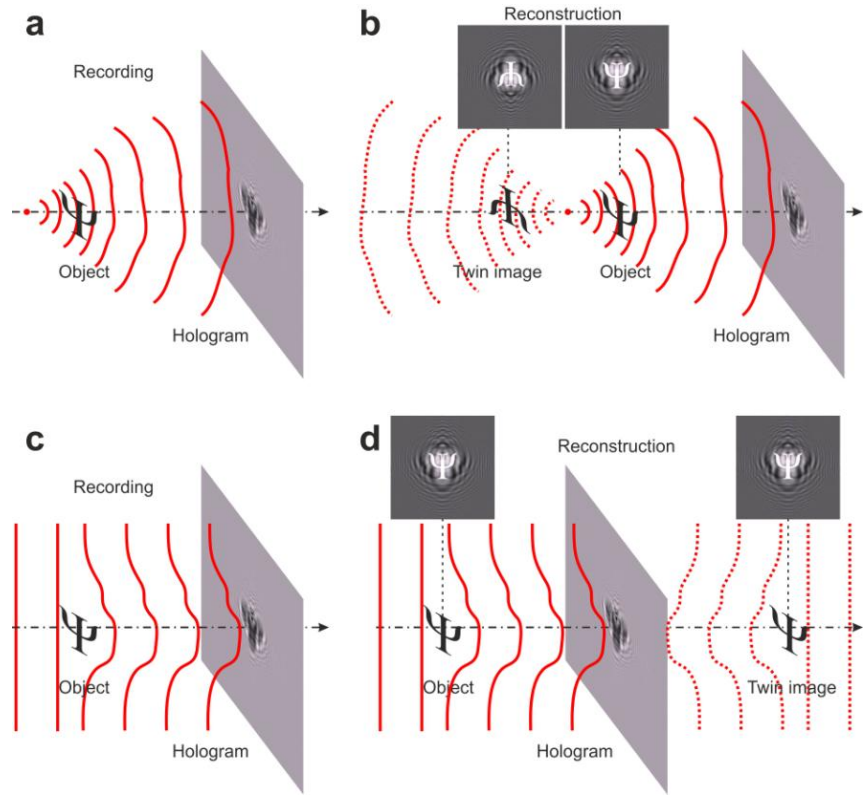

**Fig. 1**. Position of the object and its twin image during recording and reconstruction with (a–b) spherical waves and (c–d) plane waves.

The twin image problem has a number of solutions. One prominent solution is off-axis holography. Interestingly, off-axis holography was first demonstrated with electrons. In 1956 Möllenstedt and Düker invented the electron "biprism", a positively charged wire which allows splitting an electron wave into two parts, thus acting analogously to an optical prism [6]. A year later, in 1957, off-axis electron holography was first demonstrated by Möllenstedt and Keller [7], when they measured phase shift due to the electrostatic potential in carbon films. With the invention of lasers – bright coherent sources – holography became accessible. The first optical off-axis hologram was recorded and reconstructed with laser light by Leith and Upatnieks in 1963 [8]. Although off-axis holography allows elimination of twin images, in-line holography has an important advantage: it does not require any additional optical elements for splitting the beam, and therefore it is simpler to realize experimentally. Therefore, the search for alternative solutions to the twin image problem continued.

Mathematically, the holographic principle can be expressed through the following formula:

$$H(X,Y) = |R(X,Y) + O(X,Y)|^2 =$$
$$= |R(X,Y)|^2 + |O(X,Y)|^2 + R^*(X,Y)O(X,Y) + R(X,Y)O^*(X,Y). \quad (1)$$

Here $R(X,Y)$ is the reference wave, $O(X,Y)$ is the object wave, and $(X,Y)$ is the coordinate on the detector. In Eq. (1) $|R(X,Y)|^2$ is a constant term, $|O(X,Y)|^2$ is the term which is usually smaller than the other terms and can be neglected, and the sum of two terms $R^*(X,Y)O(X,Y) + R(X,Y)O^*(X,Y)$ describes the interference pattern. In the reconstruction, the hologram $H(X,Y)$ is illuminated with the reference wave $R(X,Y)$ so that the last two terms in Eq. (1) give:

$$R(X,Y)H(X,Y) \propto |R(X,Y)|^2 O(X,Y) + R^2(X,Y)O^*(X,Y). \quad (2)$$

The first term in Eq. (2) provides the object wave distribution and the second term in Eq. (2) provides the distribution of the so-called "twin image", which appears during the reconstruction along with the object reconstruction.

With the availability of computers and algorithms, the reconstruction of holograms became a numerical process [9]. If the phase distribution in the hologram plane would be known, the complete wavefront would allow twin-image-free reconstruction. However, the phase information is lost during the measurement and needs to be recovered which constitutes the so-called "phase problem". With the invention of iterative algorithms for phase retrieval [10], solutions to the twin image problem were sought by applying iterative methods. Applying iterative phase retrieval reconstruction not only eliminates the twin images, but it also allows reconstructing quantitatively correct phase and absorption distribution of the object, which is an even more important achievement than the removal of twin images [11]. For this reason, iterative phase retrieval methods became quite popular for the reconstruction of digital holograms. Moreover, currently, only iterative phase retrieval methods allow the reconstruction of phase objects from their in-line holograms [12]. Iterative phase retrieval algorithms can also be applied in off-axis holography for suppression of the so-called "zero-term" [13]. However, iterative phase retrieval algorithms are mainly applied for the reconstruction of in-line holograms, and therefore the current paper is limited to the case of in-line holography realized with waves of single wavelength. Moreover, the current paper addresses only the iterative phase retrieval algorithms that employ back-and-forth propagation of the wavefront between two planes with constraints superimposed in these two planes, as was originally proposed by Gerchberg and Saxton [10]; these are the so-called Gerchberg–Saxton (GS)-based algorithms. For further reading, the overview of phase retrieval algorithms with application to optical imaging by Shechtman et al is recommended [14].

The sections bellow are organized in the following order: introduction to the principles of iterative phase retrieval algorithms; a typical protocol of applying an iterative phase retrieval algorithm to a single-shot hologram with examples including phase objects; iterative phase retrieval algorithms applied to two or more holograms for the reconstruction 3D objects including phase objects; discussion of resolution enhancement by applying extrapolation-assisted iterative phase retrieval algorithms; discussion comparing reconstructions obtained by iterative phase retrieval from single-shot and two or more holograms.

## 2. ITERATIVE PHASE RETRIEVAL

The first iterative algorithm for the retrieval of the phase of an optical wavefront was demonstrated by Gerchberg and Saxton in 1972 [10]. The Gerchberg-Saxton (GS) algorithm constitutes a template for iterative phase retrieval algorithms in optics, and it is depicted in Fig. 2.

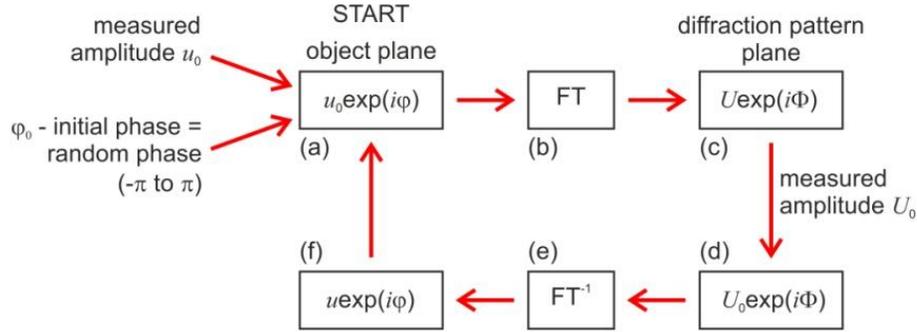

**Fig. 2**. Schematic drawing of Gerchberg-Saxton iterative phase retrieval algorithm from two intensity measurements, adapted from [10]. The measured intensities are: $|u_0|^2$ is the intensity in the object plane and $|U_0|^2$ is the intensity in the diffraction plane. From these two measured intensities, the algorithm recovers the complex-valued distributions in the object plane ($u_0 \exp(i\varphi_0)$) and in the diffraction plane ($U_0 \exp(i\Phi_0)$). (a) The algorithm starts in the object plane, where the initial complex-valued distribution is created by combining the measured amplitude distribution with the random phase distribution. (b) The Fourier transform of the object distribution gives the complex-valued distribution in the diffraction plane (c). The amplitude distribution in the diffraction plane is replaced with the measured amplitude distribution, creating an updated distribution of the complex-valued wavefront in the diffraction plane (d). An inverse Fourier transform gives the complex-valued distribution in the object plane (f). The amplitude distribution in the object plane is replaced with the measured amplitude distribution, creating an updated object distribution for the next iteration starting at (a).

The GS algorithm requires two intensity measurements: one in the sample plane and the second one in the detector plane. The algorithm assumes that the complex-valued wavefronts in the sample and the detector planes are connected through a Fourier transform (FT) with each other. The result of the algorithm is the recovered complex-valued wavefront distributions in the sample and the diffraction planes.

Iterative phase retrieval is a key component in coherent diffraction imaging (CDI), where the diffraction pattern of an object is acquired in the far-field and the object distribution is then numerically reconstructed [15-17] by applying phase retrieval algorithms [18]. In CDI the object distribution and the far-field wavefront are related to each other by FT. In holography, the two distributions are related to each other through more involved integral transformations, such as that based on the Huygens-Fresnel principle which can be calculated by the angular spectrum method, the Rayleigh-Sommerfeld formula etc [19-21]. Because in CDI the wavefront distribution in the far-field is always the FT of the object distribution, changing object-to-detector distance only changes the magnification of the diffraction pattern. In holography, the diffracted wavefront is often acquired in the Fresnel (near-field) regime, where the wavefront distribution depends on the object-to-detector distance. This property is employed for phase retrieval from a set of holograms acquired at different object-to-detector distances, as discussed below in more detail.

## 3. Iterative phase retrieval from single-shot intensity measurement (hologram)

The initial methods of iterative reconstructions from single-shot holograms were limited to pure real-valued objects (Liu et al [22]) or to objects with an exactly known shape (object support) [23]. In 2007, Latychevskaia and Fink demonstrated twin image removal by an iterative phase retrieval algorithm which was not limited to far- or near-field regimes and did not require any a priori information about the object. It employed the simple and natural constraint that the object's absorption is positive [11]. Below we present the protocol of iterative phase retrieval from a single-shot intensity measurement (hologram).

*A. Transmission function*

The transmission function describes the interaction between the incident wave and the sample. Generally, the transmission function is a complex-valued function and it is assigned to a plane:

$$t(x, y) = \exp[-a(x, y)]\exp[i\varphi(x, y)] \qquad (3)$$

where $\exp[-a(x, y)]$ is the amplitude of the transmission function, $a(x, y)$ is the function which describes the absorption properties of the sample, $\varphi(x, y)$ is the function which describes the phase added by the sample into the passing wave, and $(x, y)$ is the coordinate in the sample plane. When a wave $u_0(x, y)$, passes through an object with the transmission function $t(x, y)$, the wavefront immediately behind the sample, or the so-called "exit wave" is given by:

$$u(x, y) = u_0(x, y) t(x, y). \qquad (4)$$

For 3D samples, the sample can be split into planes and each plane can be assigned its transmission function. The wavefront propagation is then given by propagating the wave through these planes, and such approach is called "multislicing" [24].

On the other hand, the transmission function in the object plane can be written as:

$$t(x, y) = 1 + o(x, y), \qquad (5)$$

where 1 corresponds to the transmittance in the absence of the object, and $o(x, y)$ is a complex-valued function which describes the perturbation caused by the presence of the object. Writing the transmission function as $1 + o(x, y)$ helps to identify the part of the incident beam which passes the object unscattered, thus forming the reference wave. The part of the beam scattered by the object gives rise to the object wave.

*B. Hologram recording*

A wave $Au_0(x, y)$, where $A$ is a complex-valued constant, propagates toward a distant screen illuminating it with the intensity $Au_0(x, y) \rightarrow |A|^2 |R(X, Y)|^2 = B(X, Y)$, thus providing the background $B(X,Y)$. The reference wave $Au_0(x,y)$ can be of arbitrary distribution, for example, a plane wave or a spherical wave.

When an object is placed into the beam, part of the wave will interact with the object, giving rise to the object wave. The other part of the wave will go unscattered. For simplicity, we consider a 2D object in the $(x,y)$ plane, and the distribution of the transmission function in the plane $(x,y)$ is described by $t(x,y)$. Mathematically, we can write:

$$Au_0(x,y)t(x,y) = Au_0(x,y)[1+o(x,y)] \to A[R(X,Y)+O(X,Y)]. \quad (6)$$

Here the symbol $\to$ means the wavefront forward propagation to the screen plane which is described by the integrals based on the Huygens-Fresnel principle. $R(X,Y)=1$ and $R(X,Y)=\exp(ikr_s)$ for plane and spherical wave, respectively, where $r_s = (X,Y,Z)$ is the coordinate on the screen, and $Z$ is the distance from the source to the screen; $|R(X,Y)|^2 = 1.$ The total field at the screen is the sum of the propagated reference and object waves. The interference pattern on the screen can be recorded by a sensitive medium, yielding a hologram with the transmission function

$$H(X,Y) = |A|^2 |R(X,Y)+O(X,Y)|^2. \quad (7)$$

### C. Hologram normalization

The interference pattern on the screen can be recorded by a sensitive medium, yielding a hologram with the transmission function as defined by Eq. (7). Dividing the hologram image by the background image results in $H(X,Y)/B(X,Y) = |R(X,Y)+O(X,Y)|^2$, which we call the normalized hologram. The background image should be recorded with the exact same experimental conditions as the hologram, only in the absence of the object. Alternatively, an artificial background image can be generated from a hologram by smoothing it to remove all of the interference pattern. The normalized hologram is independent on $|A|^2$, where $|A|^2$ includes such factors as the point source intensity, camera sensitivity, image intensity scale defined by the image format, etc. The iterative reconstruction routine can be applied to such normalized hologram without knowing the details of the data acquisition.

### D. Iterative algorithm

A general scheme of employing iterative phase retrieval in digital holography is depicted in Fig. 3. An important requirement for applying iterative phase retrieval reconstruction to a single-shot hologram is that the sample must be located in one plane, or it should be sufficiently thin so that it can be approximated by a distribution in one plane (for example, polystyrene spheres on glass). The constraints are then applied in two planes: the sample plane and the hologram plane [11].

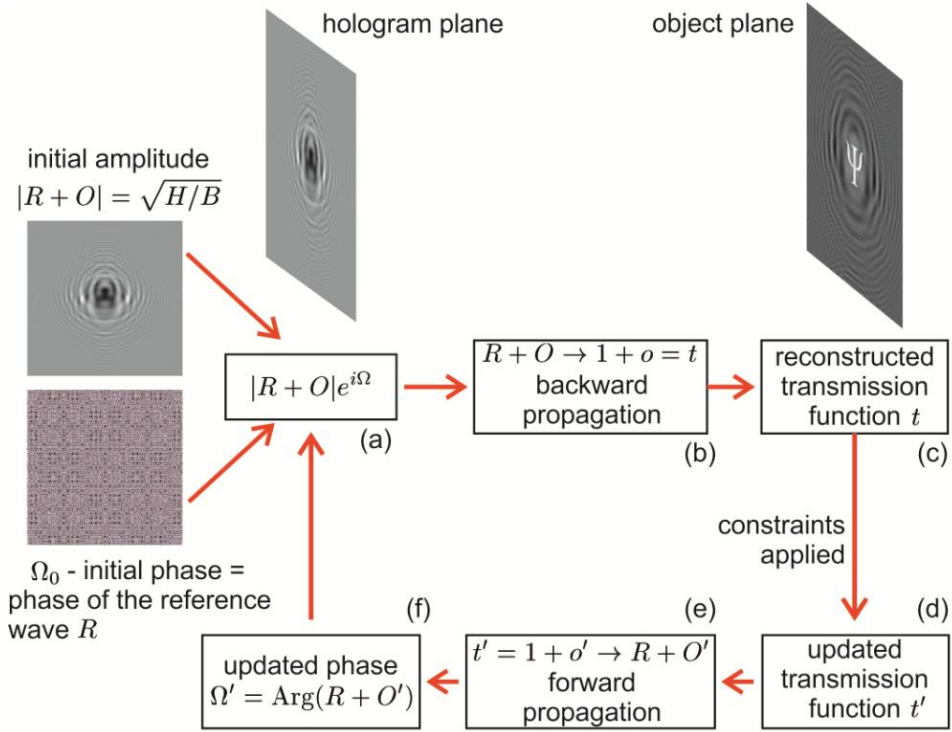

**Fig. 3**. A general scheme of iterative phase retrieval from a single-shot intensity measurement (hologram), adapted from [10]. (a) The algorithm starts in the hologram plane, where the initial complex-valued distribution is created by combining the measured amplitude distribution with the phase of the reference wave. (b) The wavefront propagates from the hologram plane to the sample plane, where it gives the distribution complex-valued transmission function $t(x,y)$. (c) Constraints in the sample plane are applied, and the transmission function $t'(x,y)$ is updated (d). (e) The wavefront is propagated from the sample plane to the detector plane (f). The amplitude of the wavefront distribution in the hologram plane is replaced with the measured amplitude. The complex-valued wavefront distribution in the detector plane is updated for the next iteration starting at (a).

*E. Constraints*

The constraint applied in the hologram plane is that the amplitude of the wavefront should be the same as the square root of the measured intensity. Therefore, at each iteration the updated amplitude in the hologram plane is replaced with the square root of the measured intensity.

Various constraints can be applied in the object plane. One possible constraint is that the absorption of the object must be positive. This does not require any a priori information about the object, since all physical objects exhibit positive absorption. To apply the positive absorption constraint, the hologram must be normalized by division with the background (as described above) so that the quantitatively correct absorption distribution can be extracted from the transmission function by applying Eq. (3) [11, 25]. If the object shape is a priori known, a support constraint in the form of a tight mask can be applied [23, 26]. If it is a priori known that the object is real-valued, the object phase during the reconstruction can be set to zero. However, in general, objects are described by complex-valued transmission functions and amplitude and phase can simultaneously be recovered. A combination of the positive absorption and finite support constraints allows faster reconstruction of samples with absorption and significant phase shift [27]. An example of such reconstruction is shown in Fig. 4.

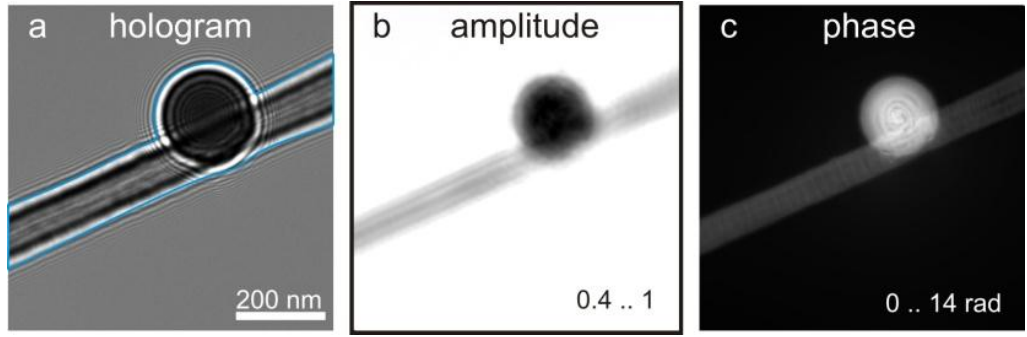

**Fig. 4**. Electron (200keV) in-line hologram of latex and its reconstruction. (a) In-line hologram of the sphere recorded at the defocus 180 μm. The blue lines mark the area outside of which the transmission was set to 1 during the iterative reconstruction. (b) Retrieved amplitude distribution of the object wave. (c) Retrieved phase distribution of the object wave. Adapted from [27].

*F. Reconstruction of phase object from its single-shot in-line hologram*

In general, a phase object cannot be reliably recovered from its single-shot in-line hologram without applying an iterative reconstruction. An example of such amplitude and phase object is illustrated in Fig. 5. Here, the transmittance and the phase in the object plane are varying in the ranges 0.6 ....1 au and 0...2 rad, respectively, as shown in Fig. 5(a). The phase of the transmitted wave in the detector plane reaches 1 rad, as shown in Fig. 5(b). Figure 5(c) exhibits the reconstructed transmittance and phase distributions. Both distributions exhibit concentric rings instead of the true object distribution. It is therefore not evident that the object is reconstructed at the correct in-focus position. This can be a problem when reconstructing an experimental hologram, where the exact in-focus position of the object is not known. Moreover, neither absorption nor phase distributions are reconstructed correctly. Figure 5(d) shows transmittance and phase distributions reconstructed by the applying iterative phase retrieval procedure. Both distributions are almost perfectly recovered. The iterative phase retrieval procedure is described in detail in [12].

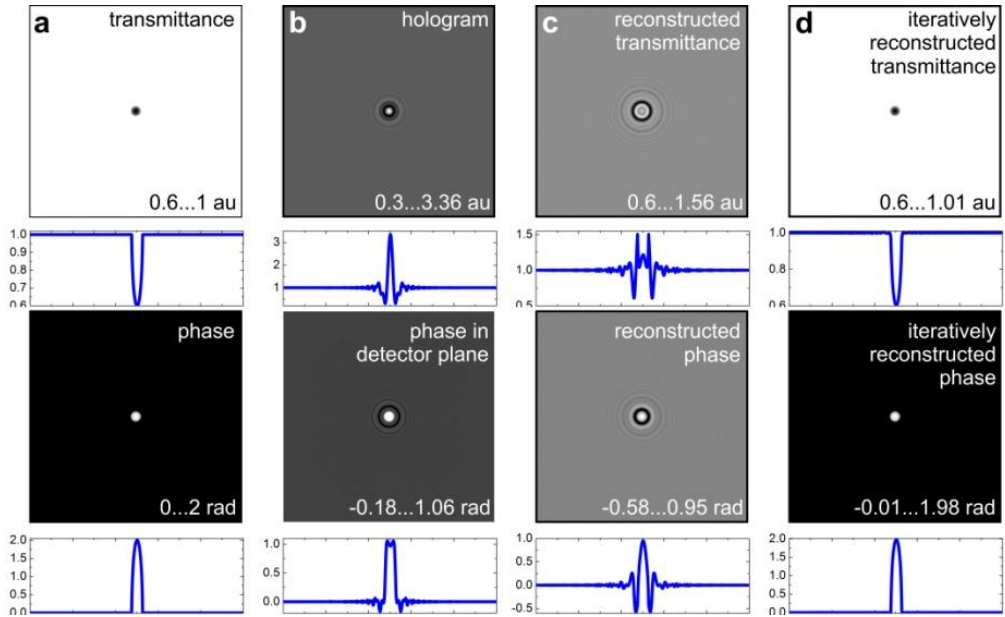

**Fig. 5**. Object with absorbing and phase-shifting properties. (a) Distributions of transmittance (top) and phase (bottom) of the object. (b) Simulated hologram (top) and phase distributions at the detector plane (bottom). (c) Reconstructed amplitude (top) and phase (bottom) distributions of the transmission function. (d) Iteratively reconstructed amplitude (top) and phase (bottom) distributions in the object plane. The blue curves are the line scans through the centers of the corresponding images. Adapted from [12].

## 4. Iterative phase retrieval from two or more intensity measurements (holograms)

The possibility of full wavefront reconstruction from a sequence of intensity measurements acquired at different object-to-detector distances was originally proposed by Schiske in 1986 for electron microscope measurements [28]. In electron holography, such approach is called "focal series reconstructions" and is successfully applied for the reconstruction of material science samples at atomic resolution [29]. In optical holography, the complete wavefront reconstruction from a sequence of intensity measurements by applying an iterative procedure was initially demonstrated in series of works in 2003 – 2006 [30-32]. The realizations of iterative phase retrieval from two or more intensity measurements have successfully demonstrated the complete wavefront recovery, even without a reference wave. However, the objects under study were limited to a 2D sample positioned at one plane [30-34]. Thus, the GS-based algorithms were successfully employed to reconstruct samples that are limited to a 2D plane; however, realistic physical objects always have some finite thickness and therefore are 3D rather than 2D objects. Here, for the first time, we demonstrate how a truly 3D object distribution, including 3D phase objects, can be recovered from two or more intensity measurements.

### A. Reconstruction of complete wavefront of 3D objects from two and more intensity measurements

Using two or more intensity measurements allows applying an iterative phase retrieval routine which calculates the propagation of the wavefront between the planes where the intensity was measured. Such approach has the advantage that absolutely no requirements are superimposed onto the sample, that is, the sample distribution can be anything. For example, the sample must not be thin or located in one plane, and, importantly, it can be 3D.

An example of a 3D sample and its reconstruction from two intensity measurements is shown in Fig. 6. In the simulations, a plane wave was assumed and the wavefront propagation is calculated by applying the angular spectrum method (ASM) [19, 35], as explained in detail elsewhere [36], here we provide the main details. In the ASM, a complex-valued wave $u_{z1}(x, y)$ at a plane $z_1$ is propagated to a plane located at $z_2$, thus giving $u_{z2}(x, y)$, by calculation of the following transformation:

$$u_{z2} = \text{FT}^{-1}\left\{\text{FT}(u_{z1})\exp\left(\frac{2\pi i \Delta z}{\lambda}\sqrt{1-\alpha^2-\beta^2}\right)\right\}, \tag{8}$$

where FT and FT$^{-1}$ are the FT and inverse FT, respectively, $(\alpha, \beta)$ are the Fourier domain coordinates, and $\Delta z = z_2 - z_1$. The FT is defined as:

$$\text{FT}(u_{z1}) = \iint u_{z1}(x, y)\exp\left[-2\pi i z\left(\frac{\alpha}{\lambda}x + \frac{\beta}{\lambda}y\right)\right]dxdy. \tag{9}$$

In the simulations here, the plane wave propagates through the sample distribution located at $z_1$, where the transmission function is described by $t_1(x_1, y_1)$. The wavefront behind the first plane is given by $u_1'(x_1, y_1) = t_1(x_1, y_1)$. Then, the resulting wavefront $u_1'(x_1, y_1)$ propagates to the next plane located at $z_2$ (propagation is calculated by ASM), where the transmission function is described by $t_2(x_2, y_2)$; the propagation distance is $(z_2 - z_1)$. The propagated wavefront in the $(x_2, y_2)$-plane is described by $u_2(x_2, y_2)$. The wavefront behind the second plane is given by $u_2'(x_2, y_2) = u_2(x_2, y_2)t_2(x_2, y_2)$, and so forth. The distances between the plane can be arbitrary, and for a dense 3D sample they can be infinitely small. After passing the last plane within the sample distribution, the exit wave is propagated toward the detector. The detector can be shifted along the z-axis, so that two or more intensity distributions are measured at different distances from the sample: $H_1, H_2,...$

The reconstruction is obtained by the GS algorithm. The wavefront propagates between the two planes $H_1$ and $H_2$ back and forth. At each iteration the phase distributions are updated and the amplitude distributions are replaced with the measured amplitudes. The initial phase distribution is zero. In the GS algorithm, knowledge of the sample plane locations, or any other information about the object is not needed. The output of the GS algorithm is two complex-valued distributions in the two hologram planes. The sample distribution is then reconstructed by backward propagation of the wavefront from one of the holograms, and different parts of the sample distribution are found in-focus at different z-planes. In the simulation, the sample consisted of four objects located at different z-distances, as shown in Fig. 6(a). Figure 6(b) depicts the reconstructions obtained by wavefront backward propagation to different z-planes within the sample distribution. It is apparent that each of four objects is correctly reconstructed twin-image free. The remaining superimposed signal is the out-of-focus signal from other objects.

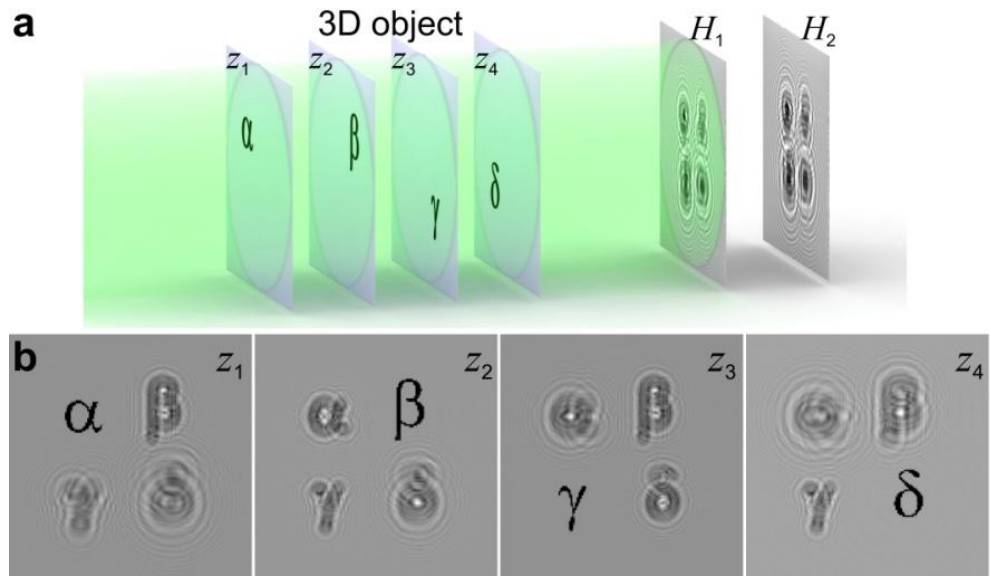

**Fig. 6**. Reconstruction of 3D objects from two or more intensity measurements. (a) Experimental arrangement. The 3D sample is represented by a set of planes at different z-positions. Here the sample is sampled with four planes. Two holograms are acquired at different distances from the sample, $H_1$ and $H_2$. (b) Reconstructed amplitude distributions at the four planes within the sample distribution. Parameters of the simulations: wavelength is 532nm, sample size is 1000 μm × 1000 μm, sampled with 1000 × 1000 pixels, distances between the planes within the sample are 50 mm, and $H_1$ and $H_2$ are acquired at distances 200 μm and 300 μm from the sample, respectively. In (b) only the central parts of the reconstructed distributions, 150 μm × 150 μm, sampled with 150 × 150 pixels, are shown.

### B. Reconstruction of complete wavefront of 3D phase objects from two or more intensity measurements

By applying iterative reconstruction from two or more measurements, a complete wavefront of 3D phase objects can be reconstructed, which are known to be difficult for reconstruction from their in-line holograms. This is possible because, as mentioned above, no constraints on the sample distribution are implied. A simulated example is shown in Fig. 7. Here, four spherical objects of 10 μm in diameter with no absorption and phase shift up to 3 radians are located at different z-distances. Two holograms are acquired at different distances from the sample (Fig. 7(a)). The reconstruction is performed by applying the GS algorithm as described above and the reconstructed sample distributions at four planes are shown in Fig. 7(b). Note that the spheres disappear in the amplitude reconstruction when at in-focus position, as can be expected. The phase values of 3 radian phase shift are correctly recovered for each sphere at its in-focus position, Fig. 7(c)-(d).

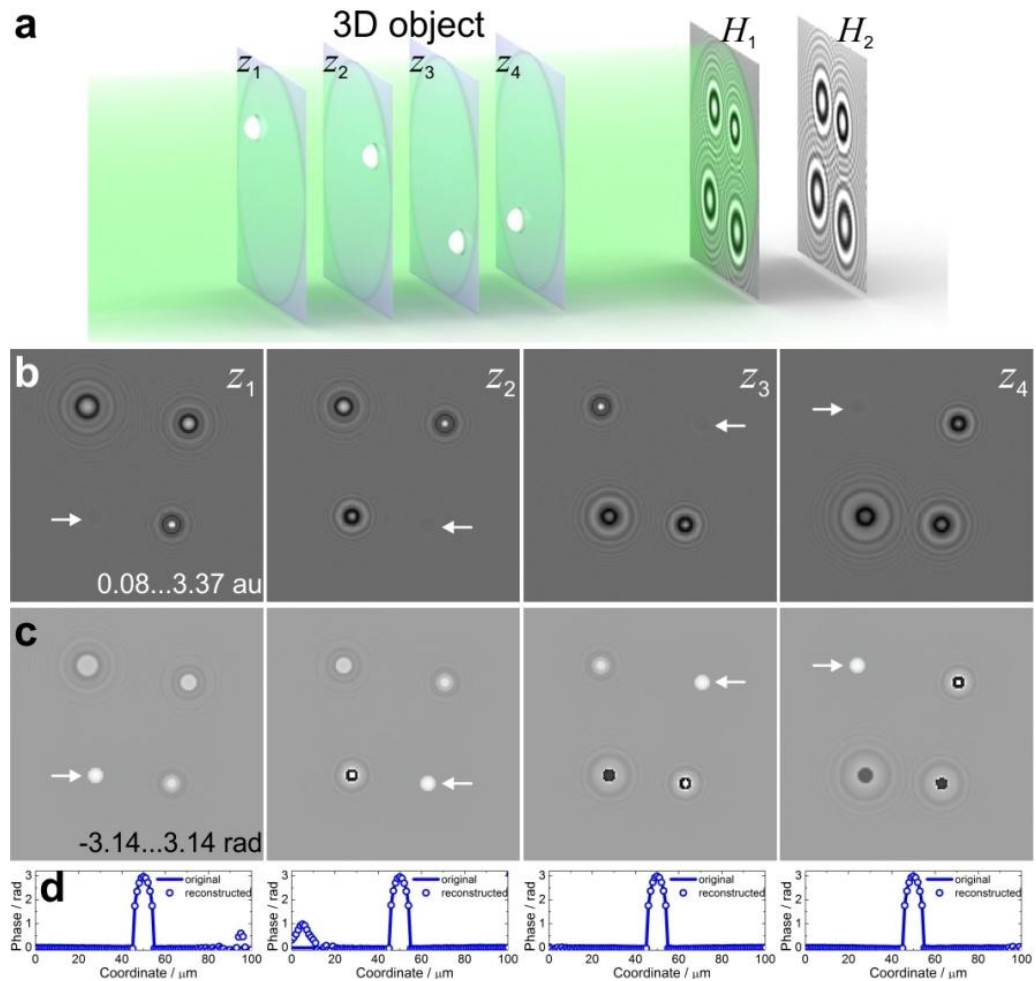

**Fig. 7**. Reconstruction of 3D phase objects from two or more intensity measurements. (a) Experimental arrangement. The 3D sample is represented by a set of planes at different z-positions. Here the sample is sampled with four planes. Two holograms are acquired at different distances from the sample, $H_1$ and $H_2$. (b) Reconstructed amplitude and (c) phase distributions at the four planes. The white arrows indicate in-focus reconstructed spheres. (d) Profiles through the reconstructed phase distributions. Parameters of the simulations are the same as in Fig. 6.

Unlike previous GS-based iterative phase retrieval algorithms that employ two or more intensity measurements, this method can be applied for thick samples. Once the complete complex-valued wavefront is reconstructed at one of the planes, it can be propagated backward, and the complex-valued exit wave can be fully recovered. The exit wave, in turn, contains all the information about all the diffraction events that took place during the wave propagation through the 3D sample. There is no restriction on thickness of the sample or on the number of diffraction events. The sample does not need to be sparse, and/or a reference wave is not required.

It must be noted that the aforementioned out-of-focus signal can severely contaminate the reconstruction in the case of thick and non-sparse samples. In this case, additional methods similar to 3D deconvolution [37], should be applied to the reconstructed wavefront to remove the out-of-focus signal.

## 5. Iterative phase retrieval with extrapolation

The resolution in digital holography is limited by the numerical aperture of the optical arrangement, in particular, by the size of the holographic record. Recently, an extrapolation method was proposed which is based on iterative phase retrieval and allows circumventing this limit by self-extrapolating conventionally acquired experimental holograms beyond the experimentally acquired area [38]. At the beginning of the algorithm, the hologram is padded by zeros (or random numbers). During the iterative reconstruction procedure, the wavefront beyond the experimentally detected area is retrieved, and the hologram reconstruction shows enhanced resolution. Moreover, if part of the hologram is missing, it can be recovered during the iterative procedure. An example is shown in Fig. 8. Here the sample exhibits four circles (Fig. 8(a)), which can be recognized in the reconstruction of their hologram (Fig. 8(b)). When only a fraction of the hologram is available (Fig. 8(c)), the reconstructed object hardly resembles the original object. When the reconstruction is obtained by the iterative procedure with extrapolation, the missing part of the hologram can be restored, and the reconstructed object appears almost matching the original distribution (Fig. 8(d)). The fact that even a fraction of a hologram is sufficient to recover the object distribution confirms the following words of Gabor: "This interference pattern I called a "hologram", from the Greek word "holos" the whole, because it contained the whole information." [39]. Iterative phase retrieval with extrapolation has already been successfully applied for resolution enhancement in terahertz holograms [40-42], where it demonstrated resolution enhancement so that features of sub-wavelength size (35 μm) were resolved from a 2.52 THz hologram (118.83 μm wavelength) [40]. Thus, the extrapolation can improve the resolution by a few times and even allows resolving of features that are smaller than the wavelength.

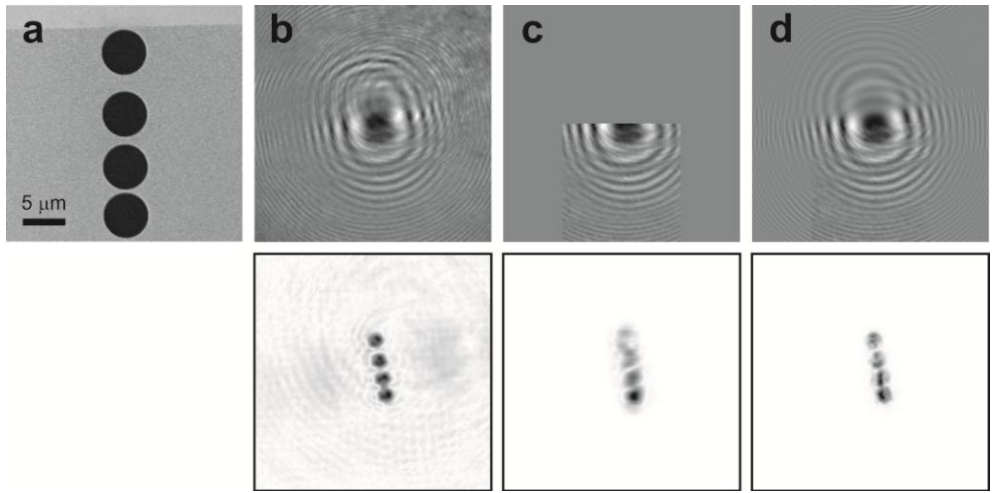

**Fig. 8.** Resolution enhancement in digital holography by self-extrapolation of a hologram. (a) Scanning electron microscope image of the sample. (b) 1000 × 1000 pixel experimental optical hologram of the sample and its reconstruction (bottom). (c) Self-extrapolation of a piece of the hologram. The selected 500 × 500 pixel part of the hologram padded with zeros up to 1000 × 1000 pixels and the corresponding reconstruction (bottom). (d) 1000 × 1000 pixel self-extrapolated hologram from (c) after 300 iterations and its reconstruction (bottom). The details of the experiment and reconstruction procedure are available in [38].

## 6. Discussion

We discussed conventional iterative phase retrieval algorithms applied in digital holography and presented a method for reconstruction of a 3D sample from two or more holograms. In discussion here, we compare the two cases of reconstruction from a single-shot hologram and from two or more holograms.

**Single-shot hologram**. Important requirement for iterative reconstruction from a single-shot hologram is that the sample must be 2D and located in one plane, or it should be sufficiently thin to be approximated by a distribution in one plane (for example, polystyrene spheres on glass). The sample can consist of phase objects. The constraints are applied in the detector and sample planes.

**Two or more holograms acquired at different distance from the sample**. In this case, there are no limitations on the sample distribution because the iterative phase retrieval is performed between the two (or more) intensity measurements outside the sample domain. Most importantly, the sample can be 3D. This method can be applied for thick biological samples.

The reason why a 3D sample can be reconstructed only from two or more intensity measurements is as follows. For a 2D sample one can set a constraint that the phase distribution in the sample plane is zero except for the phase shift introduced by the object itself. For a 3D sample, the situation is different. In any selected 2D plane crossing the 3D sample, there will be parts of the sample which are out of focus with respect to this selected plane. The wavefronts from the out-of-focus objects are spread over the selected plane, contributing nonzero amplitude and phase distribution in the selected plane. As a consequence, no specific constraints, such as zero phase or mask support, can be imposed in the selected plane. For this reason, for a truly 3D object, it is not possible to create an iterative phase retrieval algorithm from a single-shot intensity measurement.

The same argument can be re-phrased quantitatively in terms of number of equations and number of unknowns. The measured intensity distribution is given through diffraction integrals as a function of sample distribution, thus providing a system of non-linear equations. Each intensity value measured at one pixel constitutes an equation. One measured intensity distribution sampled with $N \times N$ pixels give rise to $N^2$ equations. A system of equation might have a solution if the number of equations is equal to or exceeds the number of unknowns.

For a 2D complex-valued object distribution sampled with $M \times M$ pixels, there are $2M^2$ unknowns, and a solution can exist if $M < N/\sqrt{2}$. In holography, this condition is typically fulfilled, since the area occupied by the reference wave is always larger that of the object wave [5]. In fact, some pixels in the acquired distribution can be missing, thus reducing the number of equations, and such a hologram can still be successfully reconstructed by applying iterative phase retrieval [43]. For a 2D real-valued object distribution sampled with $M \times M$ pixels, there are $M^2$ unknowns, and a solution can exist if $M < N$.

For a 3D object, there is always a non-zero phase distribution in any selected plane within the sample. This leads to the total number of unknowns (amplitude and phases) of $2N^2$, which exceeds the number of equations $N^2$, and the problem cannot have a solution. For $n$ intensity measurements there are $nN^2$ equations. In this case the condition that the number of equations exceeds the number of unknowns is fulfilled: $nN^2 > 2N^2$, at $n > 2$. Therefore, the problem of reconstructing a 3D sample from two or more intensity measurements can have a solution. In fact, two intensity measurements can be sufficient. We demonstrated a reconstruction of a 3D sample from two intensity measurements for amplitude and phase objects.